\documentstyle[12pt]{report}

\begin{document}

\begin{center}
{\bf \Large Critical holes in undercooled wetting layers}
\vspace{1cm}

G. Foltin\footnote{Present address: Lyman Laboratory of Physics, Harvard University, Cambridge, 
Massachusetts 02138}, R. Bausch, R. Blossey
\vspace{1cm}

Institut f\"ur Theoretische Physik IV, Heinrich--Heine--Universit\"at D\"usseldorf,
Universit\"atsstra{\ss}e 1, D-40225 D\"usseldorf, Germany
\vspace{1cm}

(\today)
\end{center}
\vspace{1cm}

{\bf Abstract.}\hspace{0.5cm}
The profile of a critical hole in an undercooled wetting layer is determined
by the saddle-point equation of a standard interface Hamiltonian supported by
convenient boundary conditions. It is shown that this saddle-point equation can
be mapped onto an autonomous dynamical system in a three-dimensional phase space.
The corresponding flux has a polynomial form and in general displays four fixed points,
each with different stability properties. On the basis of this picture we derive
the thermodynamic behaviour of critical holes in three different nucleation regimes
of the phase diagram.
\vspace{2cm}

\setcounter{chapter}{1}
\setcounter{equation}{0}

{\bf 1. Introduction}
\vspace{0.5cm}

The equilibrium thickness of a wetting layer on a wall is a convenient order
parameter for wetting phase transitions [1]. Fig.1 shows the phase diagram for
a first-order wetting transition in terms of temperature $ T $ and chemical potential $ \mu $ [1].
Above the coexistence value $ \mu_c $ of the two-fluid bulk system the layer thickness
is infinite whereas for $ h \equiv \mu - \mu_c < 0 $ it is finite. In the limit $ h 
\rightarrow 0 $ from below the layer thickness continuously runs to infinity above the
wetting temperature $ T_w $, but has an infinite jump across the partial-wetting line
$ T < T_w $, $ h = 0 $. A finite jump from a thin to a thick layer occurs when the prewetting
line $ T_p(h) $ is crossed from the region $ T < T_p(h) $ to $ T > T_p(h) $. This jump runs
to infinity when $ T_w $ is approached along the prewetting line and it disappears at the
prewetting critical point $ T_{pc} $.
\vspace{0.5cm}

A metastable wetting state can be generated by overheating a thin layer from $ T < T_p(h) $ into
a nucleation region bounded by $ T_p(h) $ and an upper spinodal line $ T_{us}(h) $. The 
transition to the stable thick-layer configuration occurs via the random formation of 
droplets on the thin layer and growth of the supercritical droplets [2]. Close to the 
prewetting line the critical droplets have a cylindrical shape with a diverging radius 
at $ T_p(h) $. This has been pointed out by Joanny and de Gennes [3] who have chosen the
name pancake droplets for this kind of critical nuclei.
\vspace{0.5cm}

It is also possible to undercool a thick wetting layer from $ T > T_p(h) $ into a second
nucleation region located between $ T_p(h) $ and a lower spinodal line $ T_{ls}(h) $. In this
case the critical nuclei are holes in the layer which near $ T_p(h) $ are mirror images of the
pancake droplets. Close to the partial-wetting line, however, the critical holes
have a funnel-shaped profile with a diverging depth $ F_c $ but a finite inner radius $ R_c $ at 
$ h = 0 $. There exists a third regime, adjacent to the wetting transition point $ T = T_w $, $ h = 0 $, 
at which $ F_c $ and $ R_c $ both diverge. We will refer to these regimes as the pre-dewetting,
the partial-dewetting, and the complete-dewetting regime.
\vspace{0.5cm}

In order to calculate the near-coexistence behaviour of $ F_c $, $ R_c $, and of the excess
free energy $ E_c $ of the critical hole in all regimes, we need some general information on the
critical-hole profile. This is extracted from the saddle-point equation of a standard interface
Hamiltonian [4] (which also determines the profile of critical droplets). As a general result 
this approach confirms that the size of critical nuclei diverges at the first-order lines
$ h = 0 $, $ T \leq T_w $ and $ T = T_p(h) $, whereas it shrinks to zero at the spinodal lines.
\vspace{0.5cm}

For macroscopic critical nuclei in the region $ h \approx 0 $, $ T \leq T_w $ the saddle-point equation
can be mapped onto an autonomous dynamical system in a three-dimensional 
phase space. The flux of this system is polynomial and has a surprisingly rich fixed-point structure in the
subspace of critical holes. One of these fixed points only exists, if the bulk dimension $ d $
of the system is smaller than some critical dimension $ d_1 $. In this case there appear two
different sets of physical trajectories in the flow diagram which for $ T \leq T_w $ correspond to
critical holes at $ h = 0 $, and at $ h \neq 0 $, respectively. Only the latter continue to exist
for $ d > d_1 $, and the asymptotic behaviour of their trajectories determines the $T$, $h$-dependence 
of $ F_c $, $ R_c $, and $ E_c $. For critical holes
at $ h = 0 $ the only nontrivial result of this kind is the $T$-dependence of $ R_c $.
\vspace{0.5cm}

\setcounter{chapter}{2}
\setcounter{equation}{0}

{\bf 2. Interface model for critical holes}
\vspace{0.5cm}

Our calculations are based on the Hamiltonian [4] 
\begin{equation}
H[f] = \int d^{\, d-1}x \left[\frac{\gamma}{2}(\nabla f)^2 + V(f) - hf\right]
\end{equation}
where $ f(x) $ is the local thickness of the wetting layer on the $(d-1)$-dimensional
planar wall of the system. The gradient term describes long-wavelength capillary 
excitations with a stiffness constant $ \gamma $. $ V(f) $ is an effective potential
of the form shown in fig.2, where the repulsive core simulates the wall, and $ V(f) \rightarrow 0 $
for $ f \rightarrow \infty $. The height of the minimum at $ f_{00} $ changes with temperature, and,
in a mean-field picture, $ V(f_{00}) = 0 $ at $ T = T_w $. Therefore, the spreading coefficient
$ S \equiv V(f_{00}) $ can be used to measure the temperature distance from the wetting transition
point [1].
\vspace{0.5cm}

In the full potential $ \Phi(f) \equiv V(f) - hf $ the minimum of $ V(f) $ at $ f = \infty $ is,
for $ h < 0 $, shifted to a finite value $ f_1 $, as in fig.3. Along the prewetting line the
two minima of $ \Phi(f) $ have equal height, and they coincide at the prewetting critical point. 
The situation of fig.3 corresponds to some point in the lower nucleation region of fig.1. At the
lower spinodal line $ f_1 $ has reached the local maximum in $ \Phi(f) $, changing it into a turning
point.
\vspace{0.5cm}

Under the assumption of cylinder symmetry of critical holes, their radial profile $ f(r) $ 
obeys the saddle-point equation 
\begin{equation}
\gamma f''(r) + \frac{d-2}{r}\gamma f'(r) = dV/df(r) - h.
\end{equation}
Convenient boundary conditions for such objects are $ f'(r=0) = 0 $ and $ f(r = \infty) = f_1 $
for $ h < 0 $ or $ f'(r=\infty) = 0 $ for $ h = 0 $, respectively. Nontrivial solutions of the
latter type can in fact exist in a certain range of dimensions, as discussed below. 
\vspace{0.5cm}

In order 
to prove the existence of critical holes inside the nucleation region at $ h < 0 $, we adopt
an argument by Coleman [5]. He considered (2.2) as an equation of motion for a fictitious particle
with position $ f $ moving in time $ r $ in a potential $ - \Phi(f) $ in presence of a time-dependent
friction term. At time $ r = 0 $ the particle has to start with zero velocity from a position $ f_0 $ 
such that asymptotically for $ r \rightarrow \infty $ it comes to rest on top of the hill at $ f_1 $,
generating e.g. the critical-hole profile of fig.4. The existence of this marginal solution is implied
by continuity from that of undershooting and overshooting solutions. Undershooting solutions obviously
are produced by choosing a starting position sufficiently close to the local minimum in $ - \Phi(f) $
where the particle then eventually comes to rest. On the other hand, by choosing the starting position
sufficiently close to the higher maximum in $ - \Phi(f) $, the particle spends an arbitarily long time
to reach a position $ f^* < f_1 $, where $ \Phi(f^*) = \Phi(f_1) $. This allows to neglect the friction
term in the particle motion for $ f > f^*$ so that by energy conservation the particle will overshoot 
the hill at $ f_1 $.
\vspace{0.5cm}

Near interior points of the prewetting line $ \Phi(f) $ has a double-well form which for the radial
profile of a critical hole leads to a kink solution. In the limit $ T \rightarrow T_p(h) $ the position
$ R_c $ of the turning point of the kink runs to infinity, resembling the behaviour of a pancake droplet.
\vspace{0.5cm}

When some point on the line $ h = 0 $, $ T \leq T_w $ is approached, $ f_1 $ and consequently the critical
depth $ F_c \equiv f_1 - f_0 $ diverges. In this regime the critical-hole profile at macroscopic distances
from the wall is determined by the asymptotic behaviour of $ V(f) $ for $ f \rightarrow \infty $.
For long-range molecular interactions this reads
\begin{equation}
V(f) = Af^{\, 1 - \sigma},\,\,\,\,\, {\mbox{ $ for $ }} \sigma > 1
\end{equation}
where $ A $ is the Hamaker constant, and $ \sigma = 3 $ or $ \sigma = 4 $ for non-retarded or
retarded van der Waals forces, respectively [1]. By extrapolation the macroscopic profile $ F(r) $
of a critical hole can then be defined as the solution of the differential equation
\begin{equation}
\gamma F''(r) + \frac{d-2}{r}\gamma F'(r) = - A (\sigma - 1)F^{\, - \sigma} - h
\end{equation}
with the new boundary conditions $ F(r=R_c) = 0 $ and $ F(r=\infty) = F_1 $ at $ h < 0 $ or
$ F'(r=\infty) = 0 $ at $ h = 0 $, respectively. Undershooting and overshooting solutions can still
be created, now by controlling the initial velocity $ F'(r=R_c) $. 
\vspace{0.5cm}

Since for $ r \rightarrow R_c $ the friction term and
the field $ h $ can be neglected in (2.4), we find the result 
\begin{equation}
F(r) = \left[\frac{A}{2\gamma}(\sigma+1)^2 \right]^{\frac{1}{\sigma+1}} (r - R_c)^{\, \frac{2}{\sigma+1}}
\end{equation} 
which is asymptotically valid for all $ h \leq 0 $.  
\vspace{0.5cm}

For $ r \rightarrow \infty $ and $ h \neq 0 $ a linear expansion of (2.4) in $ F_c - F(r) $ leads to
a Bessel-type differential equation. This implies the asymptotic form 
\begin{equation}
F(r) = F_c\left[1 - C\left(\frac{r}{R^*}\right)^{\, \frac{2-d}{2}} e^{\, -r/R^*} \right]
\end{equation} 
where $ R^* \equiv \left[A\sigma(\sigma-1)/\gamma\right]^{\, -1/2} F_c^{\, (\sigma + 1)/2} $, and
$ C $ is an integration constant. If, as an approximation to the full solution of (2.4), the expressions
(2.5) and (2.6) and their derivatives are matched at some value $ r = R_m $, it turns out that 
$ R_m \sim R^* $ and $ C $ is of order 1. 
\vspace{0.5cm}

For $ r \rightarrow \infty $ and $ h = 0 $ the left-hand side in eq. (2.4) dominates for
$ d < d_1(\sigma) $ where 
\begin{equation}
d_1(\sigma) \equiv \frac{3\sigma + 1}{\sigma + 1}\,\,\, ,
\end{equation} 
and leads to the behaviour,
\begin{equation}
F(r) = F^* D \left(\frac{r}{R_c}\right)^{\, 3 - d}\,\,\, .
\end{equation} 
Here the amplitude $ F^* \equiv \left[A(\sigma + 1)^2/8\gamma\right]^{\, \frac{1}{\sigma + 1}}
R_c^{\, \frac{2}{\sigma+1}} $ has been adopted from the previously derived [6] exact solution
\begin{equation}
\left(\frac{r}{R_c}\right)^2 - \left(\frac{F}{F^*}\right)^{\, \frac{\sigma + 1}{2}} = 1
\end{equation} 
of (2.4) at $ h = 0 $ in the dimension 
\begin{equation}
d_0(\sigma) \equiv \frac{3\sigma - 1}{\sigma + 1}\,\, ,
\end{equation} 
so that $ D = 1 $ in $ d = d_0(\sigma) $. Due to the boundary condition $ F'(r=\infty) = 0 $ the
asymptotic form (2.8) implies that critical holes at $ h = 0 $ only exist in dimensions $ d > 2 $.
The necessity of the previously mentioned additional condition $ d < d_1(\sigma) $ will become clear
through the following analysis.
\vspace{0.5cm}

\setcounter{chapter}{3}
\setcounter{equation}{0}

{\bf 3. Mapping to a dynamical system}
\vspace{0.5cm}

We now define the dimensionless quantities
\begin{eqnarray}
X & \equiv & \frac{r F'(r)}{F(r)}\, , \nonumber \\
Y & \equiv & \frac{\sigma^2 - 1}{2} \frac{A}{\gamma}
\frac{r^2}{F^{\,\sigma+1}(r)}\,\,\, , \\
Z & \equiv & -\frac{1}{2} \frac{h}{\gamma} \frac{r^2}{F(r)}\, , \nonumber
\end{eqnarray}
and consider their dependence on the time-like variable
\begin{equation}
t \equiv \ln \frac{r}{r_1} 
\end{equation}
where $ r_1 $ is an arbitrary reference scale. 
Due to (2.4) we find the set of differential equations 
\begin{eqnarray}
\dot{X} & = & (3-d)X - X^2 - \frac{2}{\sigma+1}Y + 2Z , \nonumber \\
\dot{Y} & = & 2Y(1 - \frac{\sigma+1}{2}X)\, , \\
\dot{Z} & = & 2Z(1 - \frac{1}{2}X)\, , \nonumber
\end{eqnarray}
which has the four fixed points 
\begin{eqnarray}
X_0 & = & Y_0 \, = \,Z_0 \,= \,0\, , \nonumber \\
X_1 & = & \frac{2}{\sigma + 1}\,\, , \,\,\, Y_1 \,\, = \,\, d_1(\sigma) - d\,\,\, ,\,\,\, Z_1 \, = \, 0\,\, ,
\nonumber \\
\\
X_2 & = & 3 - d\,\,\, , \,\,\, Y_2 \,\, = \,\, Z_2 \,\, = \, 0\,\, , \nonumber \\ 
X_3 & = & 2\,\,\, , \,\,\,\,\, Y_3 \, = \, 0 \,\,\,\, ,\,\, Z_3 \,\, = \,\, d - 1\, . \nonumber  
\end{eqnarray}
For $ 1 < d < d_1(\sigma) $ the fixed points (3.4) are all located in the subspace $ X \geq 0 $, $ Y \geq 0 $,
$ Z \geq 0 $, where the critical-hole trajectories occur (whereas $ F'(r) \leq 0 $ implies
$ X \leq 0 $ for critical droplets). The subscripts of the fixed-point coordinates indicate
the numbers of attractive principal directions of each of these points. 
\vspace{0.5cm}

In the plane $ Z = 0 $ the fixed point
$ P_1 $ in fig.5 attracts the physical trajectories coming from $ X = Y = \infty $ which then either
run to the droplet region $ X < 0 $ or to the more stable fixed point $ P_2 $. The first possibility
corresponds to undershooting solutions of the saddle-point equation whereas the second one 
describes solutions obeying the boundary conditions for critical holes at $ h = 0 $. 
In fact, the fixed-point value $ X_2 = 3 - d $ in connection with
the definition (3.1) of $ X $ reproduces the asymptotic behaviour (2.8) up to an undetermined
prefactor. In the limit $ d \rightarrow d_1(\sigma) $ the fixed point $ P_1 $ merges into $ P_2 $,
and in fig.5 the right section of the basin of attraction of $ P_2 $ collapses to zero, so that critical
holes at $ h = 0 $ no longer exist for $ d > d_1(\sigma) $.
\vspace{0.5cm}

If $ h < 0 $, the physical trajectories approach $ P_1 $ from $ X = Y = Z = \infty $ but now have three 
options to continue. Most of them either run into the droplet region or to the most stable fixed point $ P_3 $,
representing respectively undershooting and overshooting saddle-point solutions where the latter behave as 
$ F(r) = - h r^2/\left[2(d-1)\gamma\right] $ for $ r \rightarrow \infty $. 
The basins of attraction for these two sets of trajectories
are separated by a surface which is the support of the critical-hole trajectories.
\vspace{0.5cm}

For $ d < d_1 $ the trajectories for critical holes at $ h < 0 $ have, 
contrary to our previous belief [7], no chance to come close to
the fixed point $ P_2 $ (which led to the erroneous result (9) in [7]). This is a consequence
of the fact that, in agreement with (2.6), these trajectories for $ t \rightarrow \infty $ 
have to run to $ X = 0 $, $ Y = Z = \infty $. According to (3.3) they must, however, penetrate the
plane $ X = 2/(\sigma+1) $ above the line $ Z = (2/(\sigma + 1))(Y - Y_1) $,
and for $ X \geq 2/(\sigma+1) $ obey the condition $ \dot{Y} \leq 0 $. This is incompatible
with a visit of the fixed point $ P_2 $ which consequently is supposed to have essentially 
no influence on the critical-hole profile for $ h < 0 $.
\vspace{0.5cm}

\setcounter{chapter}{4}
\setcounter{equation}{0}

{\bf 4. Critical holes at bulk coexistence}
\vspace{0.5cm}

At bulk coexistence $ h = 0 $ there appears an infinite set of flow lines in the $X$,$Y$-plane running from
$X=Y=\infty$ to the fixed point $P_2$. This means that the saddle-point equation (2.4) for $h=0$ has
infinitely many solutions which obey the boundary conditions for critical holes. Only one of these solutions
will, however, correspond to a true minimum in the variational principle $ \delta H = 0 $.
\vspace{0.5cm}

The situation can most easily be analyzed in the special dimension $ d = d_0(\sigma) $. Then, in terms
of the variables
\begin{equation}
\eta(t) \equiv \left(\frac{r}{R_c}\right)^{\, -\frac{2}{\sigma+1}}F(r)\,\,\,, \,\,\, 
t \equiv \ln\frac{r}{R_c}\,\,,
\end{equation} 
the saddle-point equation (2.4) assumes the form [6]
\begin{equation}
{\ddot \eta} = - \frac{\partial}{\partial \eta}\left[\frac{A}{\gamma}\,R_c^2\, \eta^{\, 1 - \sigma} + 
\frac{2}{(\sigma+1)^2} \, \eta^2 \right]\, .
\end{equation}
This again can be considered as an equation of motion for a fictitious particle, now without a friction
term. As a consequence the particle energy
\begin{equation}
\varepsilon \equiv  \frac{1}{2}{\dot\eta}^2 - \frac{A}{\gamma}\, R_c^2\, \eta^{\, 1 - \sigma} - 
\frac{2}{(\sigma+1)^2} \eta^2
\end{equation}
is conserved.
\vspace{0.5cm}

Eq. (4.3) can be rewritten as 
\begin{equation}
(x-1)^2 = \frac{2}{\sigma-1}\, y - \lambda \, \frac{\sigma+1}{\sigma-1}\, y^{\, \frac{2}{\sigma+1}} + 1
\end{equation}
where
\begin{equation}
x \equiv \frac{X}{X_1}\,\,\,,\,\,\, y \equiv \frac{Y}{Y_1}\,\,\,, \,\,\, 
\lambda \equiv - \frac{\varepsilon}{\varepsilon_0}\, ,
\end{equation}
and $ \varepsilon_0 \equiv \left[2/(\sigma^2 -1)\right]
\left[(\sigma-1)(\sigma+1)^2 A R_c^2/(4\gamma)\right]^{\, 2/(\sigma+1)} $ is the maximum value of the
potential energy in (4.3). With $ \lambda $ taken as a parameter (4.4) analytically describes the full 
flow pattern of the system which is depicted in fig.6. 
Obviously this pattern is symmetric with respect to the line $ x = 1 $. For
$ x = y = 1 $ one obtains the parameter value $ \lambda = 1 $ which consequently belongs to the separatrix
running through the fixed point $ P_1 $ (see fig.5). This corresponds to zero kinetic energy of the
fictitious particle when it passes the maximum of the potential in (4.3). Values $ \lambda > 1 $ accordingly
belong to undershooting solutions whereas for $ \lambda < 1 $ one finds an infinite set of solutions 
obeying the critical-hole boundary conditions. The profile (2.9) leads to the value $ \lambda = 0 $,
i.e. a simple parabola for the corresponding flow line.
\vspace{0.5cm}
 
For $ d = d_0(\sigma) $ the energy functional $ H[F(r)] $ can be written in the form $ H = \Omega_{\, d - 1} 
\left[(\sigma^2 - 1) A Y_1^{\, \sigma}/(2\gamma)\right]^{\, 2/(\sigma+1)} u[y] $ where
\begin{equation}
u = \int_0^\infty dy \, \frac{y^{\, -\frac{\sigma+3}{\sigma+1}}}{2(x-1)} \,
\left[\frac{1}{2}x^2 + \frac{1}{\sigma - 1}y \right]\, .
\end{equation}
This implies
\begin{equation}
\frac{\delta u}{\delta x(y)} = \frac{1}{4} y^{\, - \frac{\sigma+3}{\sigma+1}}
\frac{1}{(x-1)^2}\left[x^2 - 2x - \frac{2}{\sigma-1}y \right]\, ,
\end{equation}
which shows that the variational principle $ \delta u = 0 $ leads to (4.4) with $ \lambda = 0 $. 
The special solution (2.9) therefore, in fact, corresponds to a true minimum of $ H $.
\vspace{0.5cm}

\setcounter{chapter}{5}
\setcounter{equation}{0}

{\bf 5. Scaling behaviour of critical holes}
\vspace{0.5cm}

In order to calculate the quantities $ F_c $, $ R_c $, and $ E_c $ for a critical hole at arbitrary values
of $ h $, we use the definitions
\begin{equation}
\Phi'(F_c + f_0) = 0\, ,\,\,\, E_c \equiv H[f(r)] - H[F_c + f_0]\,\, ,
\end{equation}
and extract $ R_c $ from the relation
\begin{equation}
\Phi(F_c + f_0) - \Phi(f_0) = (d-2)\gamma\int_{0}^{\infty} dr \frac{(f'(r))^2}{r}\,\, ,
\end{equation}
implied by the saddle-point equation (2.2). The second eq.(5.1) can slightly be simplified 
by use of the virial theorem which states that the potential-energy part in (5.1) is
$ (3-d)/(d-1) $ times the kinetic part. This follows from the scaling property
$ \partial H[f(\alpha r)]/\partial \alpha|_{\alpha = 1} = 0 $ which in turn is implied by
the variational principle $ \delta H[f_{\alpha}(r)] = 0 $ for the special set of
functions $ f_{\alpha}(r) \equiv f(\alpha r) $. As a result of this procedure we eventually
find 
\begin{equation}
E_c = \frac{1}{d-1} \Omega_{\, d-1} \gamma \int_{0}^{\infty} dr r^{\, d - 2} (f')^2
\end{equation}
where $ \Omega_{\, d-1} $ is the volume of the $(d-1)$-dimensional unit sphere. 
\vspace{0.5cm}

Close to the line $ h = 0 $, $ S \leq 0 $, we can in (5.1)-(5.3) neglect the microscopic
increment $ f_0 $ to $ F_c $, replace $ V(f) $ by its asymptotic form (2.3), and insert for
$ f(r) $ the macroscopic profile $ F(r) $. This leads to the result 
\begin{equation}
F_c = \frac{1}{(\sigma - 1)A}|h|^{\, -\frac{1}{\sigma}}
\end{equation}
for $ h \rightarrow 0 $, and, in leading order, to the equations
\begin{equation}
AF_c^{\, 1 - \sigma} - hF_c - S = (d-2)\gamma\int_{R_c + r_0}^{\infty} dr r^{\, -1} (F')^2(r)\,\, ,
\end{equation}
\vspace{0.2cm}
\begin{equation}
E_c = \frac{1}{d-1} \Omega_{\, d-1} \gamma \int_{R_c + r_0}^{\infty} dr r^{\, d-2} (F')^2(r)\,\, .
\end{equation}

Here a cut-off length $ r_0 \ll R_c $ has been introduced in order to cure the
artificial singularity which due to the extrapolation (2.5) occurs at $ r = R_c $ 
in the case $ \sigma \geq 3 $. 
\vspace*{0.05cm}

In (5.5) and (5.6) we now split off integrals running from $ R_c + r_0 $ to 
$(1 + \lambda)R_c $ wherein the choice $ 0 < \lambda \ll 1 $ allows to use (2.5).
In the remaining integrals we transform to the scaled variables $ r/R^* $ for
$ h < 0 $ and $ r/R_c $ for $ h = 0 $, suggested by the asymptotic forms (2.6)
and (2.8). This leads to a power in $ R^* $ and $ R_c $, respectively, where the 
cofactors are assumed to be finite in the limit $ h \rightarrow 0 $ and
$ S \rightarrow 0 $. The latter assumption is supported by the analysis of chapter 3
which leads to expect that no further singularities will occur in addition to those
implied by (2.5)-(2.8).
\vspace{0.5cm}

On a path $ S = const. $ in the partial-dewetting regime the procedure just described
leads for $ |h| \rightarrow 0 $ to a constant value of $ R_c $, and to
\begin{equation}
E_c  \sim  F_c^{\, \frac{\sigma+1}{2}(d - d_0(\sigma))} 
\end{equation}
with $ F_c $ given by (5.4). At $ S = 0 $, i.e. in the complete-dewetting regime,
we find the behaviour 
\begin{eqnarray}
R_c & \sim & F_c^{\, \frac{\sigma+1}{2}}\,\,\,\,\,\,\,\,\,\,\,\,{\mbox{ $ for $ }}\,\,\,\,\sigma < 3\,\, , 
\nonumber \\
R_c & \sim & F_c^2\ln F_c\,\,\,\,\,{\mbox{ $ for $ }}\,\,\,\,\,\sigma = 3\,\,\, ,\\
R_c & \sim & F_c^{\, \sigma - 1}\,\,\,\,\,\,\,\,\,\,\,\,{\mbox{ $ for $ }}\,\,\,\,\sigma > 1\, \nonumber
\end{eqnarray}
where again $ F_c $ has the form (5.4). Moreover, in this regime we obtain
\begin{eqnarray}
E_c & \sim & R_c^{\, d - d_0(\sigma)}\,\,\,\,\,\,\,\,\,\,{\mbox{ $ for $ }}\,\,\,\,\sigma < 3\, , \nonumber \\
E_c & \sim & R_c^{\, d-2}\ln R_c\,\,\,\,{\mbox{ $ for $ }} \,\,\,\,\,\sigma = 3\, ,  \\
E_c & \sim & R_c^{\, d-2}\,\,\,\,\,\,\,\,\,\,\,\,\,\,\,\,\,{\mbox{ $ for $ }} \,\,\,\,\,\,\,\sigma > 3\, . 
\nonumber
\end{eqnarray}
For critical holes at $ h = 0 $ the only nontrivial result is the behaviour
\begin{eqnarray}
R_c & \sim & |S|^{\, -\frac{\sigma + 1}{2(\sigma - 1)}}\,\,\,\,\,\,\,{\mbox{ $ for $ }} \,\,\,\sigma < 3\, ,  
\nonumber  \\
R_c & \sim & |S|^{\, -1}\ln |S|\,\,\,\,{\mbox{ $ for $ }} \,\,\,\sigma = 3\,\, ,  \\
R_c & \sim & |S|^{\, -1}\,\,\,\,\,\,\,\,\,\,\,\,\,\,\,\,\,\,{\mbox{ $ for $ }}\,\,\sigma > 3\,\,\, \nonumber
\end{eqnarray}
for $ |S| \rightarrow 0 $. 
\vspace{0.5cm}

In the pre-dewetting regime the asymptotic behaviour of the pancake critical holes will,
with growing distance from $ T_w $, increasingly depend on the microscopic details of the
potential $ \Phi(f) $. We therefore have to go back to the relations (5.2) and (5.3), in which we then
use the fact that $ f'(r) $ is sharply peaked at the value $ r = R_c $. This leads for any path
$ S = const. $ to a constant value of $ F_c $ at the prewetting line $ h = h_p(T) $, and to the relations
\begin{equation}
R_c \sim (h_p(T) - h)^{\, -1}\,\, ,
\end{equation}
\begin{equation}
E_c \sim (h_p(T) - h)^{\, 2-d}
\end{equation}
which are identical to those for pancake droplets [3]. When the wetting transition point is approached
along the prewetting line, $ F_c $ diverges as in (5.4).
\vspace{0.5cm}

The crossover lines, separating the pre-dewetting and the partial-dewetting regime 
from the intervening complete-dewetting 
regime (see fig.1), are of the form $ |h| \sim |S|^{\, \frac{\sigma}{\sigma - 1}}$. This is implied 
by eq. (5.5) through which the spreading coefficient $S$ enters the calculation in the form
$ S + const |h|^{\, \frac{\sigma-1}{\sigma}} $. 
\vspace{0.5cm}

Some of the results (5.4),(5.7)-(5.12) should be accessible to experiments, although (5.10) strictly
applies only to $ d < d_1(\sigma) $. For general $ d $ the behaviour (5.10) might, however, be observable
on a path $ h = const_{\,\,\, \sim}^{\,\,\, <}\,\,\, 0 $ in the partial-dewetting regime. The relevance
of the behaviour (5.7) has been pointed out previously [8] in connection with the observed anomalously long
lifetimes of undercooled wetting layers [9]. Direct observations of the nucleation of holes in wetting
layers have also been reported in the recent literature [10]. 
\vspace{0.5cm}

{\bf Acknowledgement.} We are grateful to R. Burghaus for many helpful discussions. This work has been
supported by the DFG under SFB 237 ``Unordnung und Gro{\ss}e Fluktuationen''.

\newpage
\vspace*{-2cm}

{\bf References}
\vspace{0.5cm}

[1] P.G. de Gennes, Rev. Mod. Phys. {\bf 57}, 825 (1985); S. Dietrich, in {\it Phase Transitions
and Critical Phenomena}, eds. C. Domb and J.L. Lebowitz, vol. 12 (Academic Press, London, 1988).

[2] R. Bausch, R. Blossey, Europhys. Lett. {\bf 14}, 125 (1991); Phys. Rev. {\bf E48}, 1131 (1993);

[3] J.F. Joanny, P.G. de Gennes, C.R.Acad.Sc.Paris, {\bf 303}II, 337 (1986);

[4] E. Br\'ezin, B.I. Halperin and S. Leibler, J. Phys. (Paris) {\bf 44}, 775 (1983); R. Lipowsky, D.M. Kroll
and R.K.P. Zia, Phys. Rev. {\bf B27}, 4499 (1983);

[5] S. Coleman, in {\it Aspects of Symmetry}, Cambridge University Press, Cambridge (1985);

[6] M. Burschka, R. Blossey and R. Bausch, J. Phys. {\bf A26}, L1125 (1993); R. Bausch, R. Blossey,
M.A. Burschka, J. Phys. {\bf A27}, A1405 (1994);

[7] R. Bausch, R. Blossey, G. Foltin, Physica {\bf A224}, 93 (1996);

[8] R. Bausch, R. Blossey, Phys. Rev. {\bf E50}, R1759 (1994);

[9] J.E. Rutledge, P. Taborek, Phys. Rev. Lett. {\bf 69}, 937 (1992); D. Bonn, H. Kellay and
J. Meunier, Phys. Rev. Lett. {\bf 73}, 3560 (1994);

[10] see, e.g., the review by R. Blossey, Int. J. Mod. Phys. {\bf B9}, 3489 (1995), and more recently 
P. Lambooy, K.C. Phelan, O. Haugg, and G. Krausch, Phys. Rev. Lett. {\bf 76}, 1110 (1996),
J. Bishof, D. Scherer, S. Herminghaus, and P. Leiderer, Phys. Rev. Lett. {\bf 77}, 1536 (1996).

\newpage
{\bf Figure Captions}
\vspace{1cm}

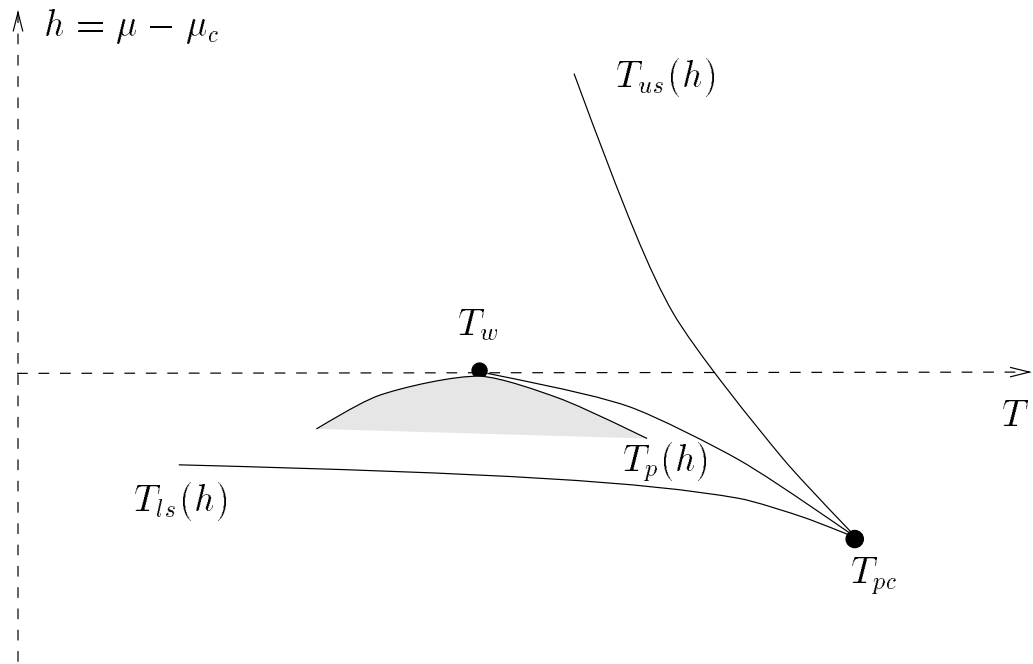
Figure 1: $T$,$\, \mu$-phase diagram for a first-order wetting transition. All symbols are
explained in the introduction. The shaded area is the complete-dewetting regime, separated
by crossover lines from the partial- and pre-dewetting regimes.  
\vspace{0.5cm}

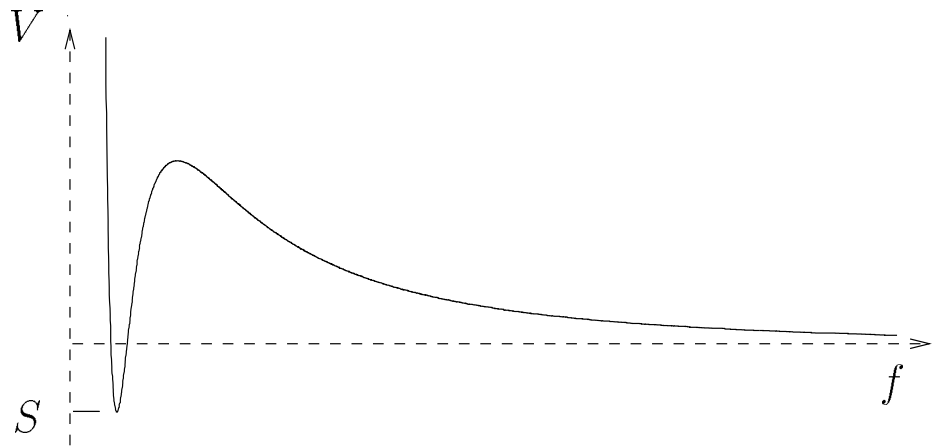
Figure 2: The effective interface potential $ V(f) $, where $ S \equiv V(f_{00}) $.
\vspace{0.5cm}

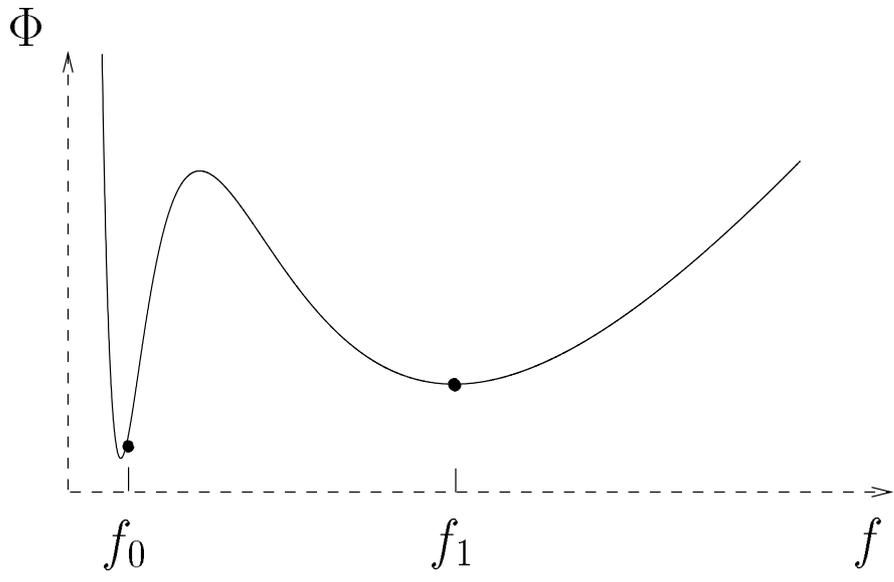
Figure 3: The full potential $ \Phi(f) \equiv V(f) - hf $. Here, $ f_1 $ is the equilibrium
thickness of the undercooled layer, and $ f_1 - f_0 $ is the depth of the critical hole.
\vspace{0.5cm}

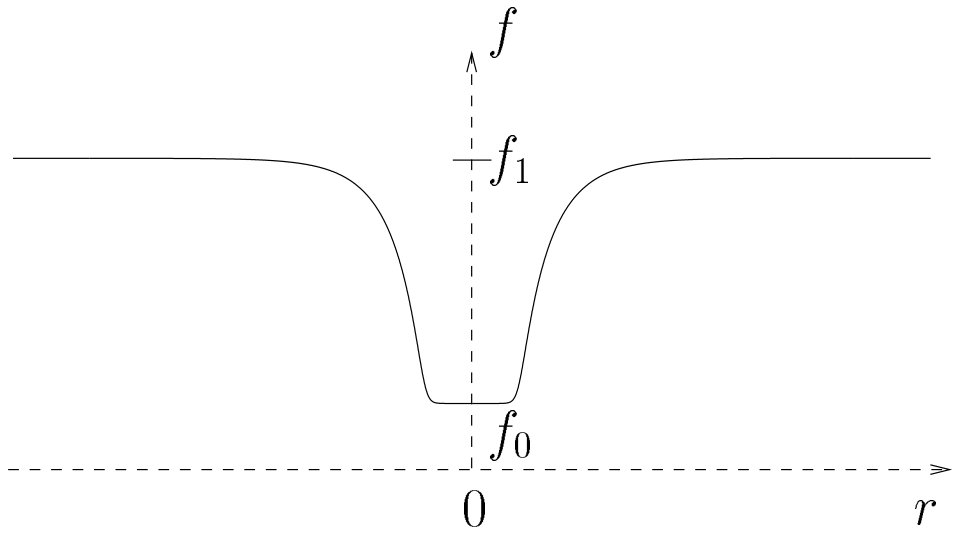
Figure 4: The profile of the critical hole corresponding to the potential in fig.4.
\vspace{0.5cm}

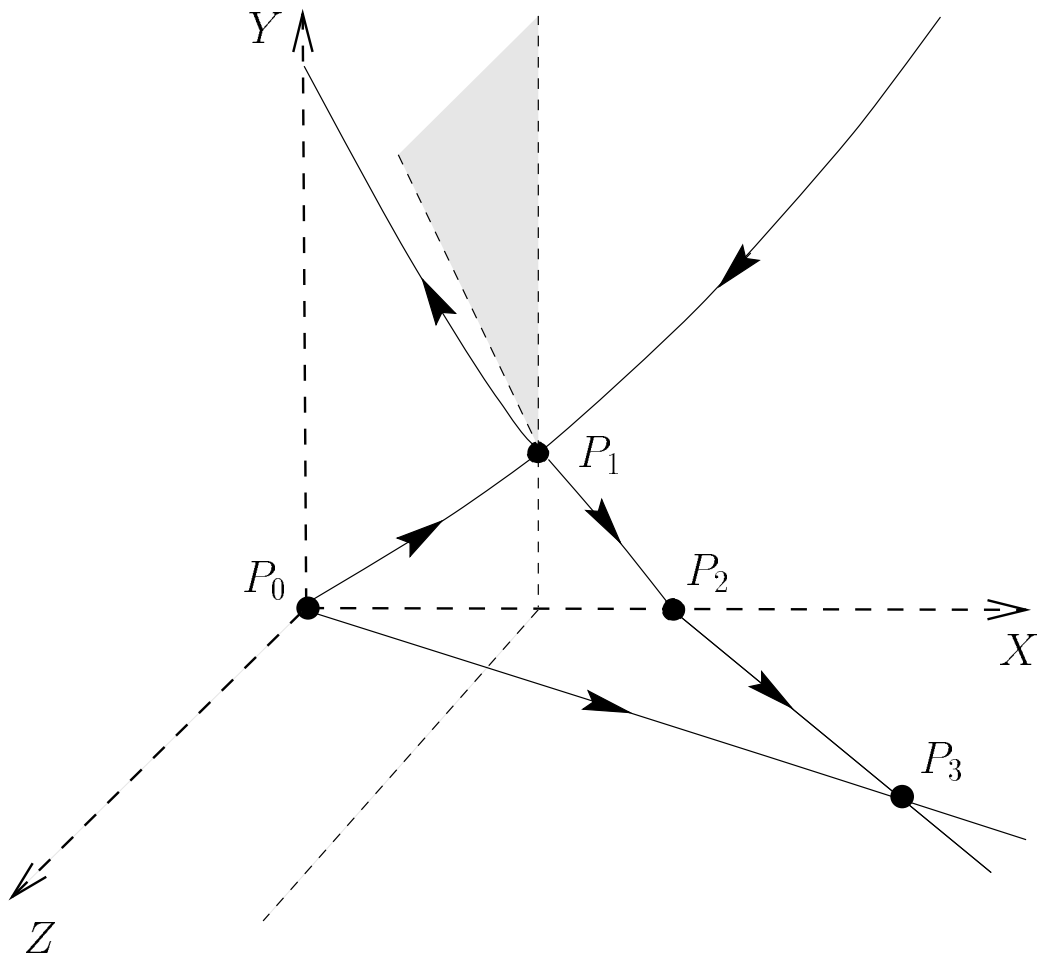
Figure 5: The flow diagram of the dynamical system (3.3) with all fixed points and their
principal directions. The shaded region is the sector in the plane $ X = 2/(\sigma + 1) $ penetrated
by the physical trajectories for $ h < 0 $.    
\vspace{0.5cm}

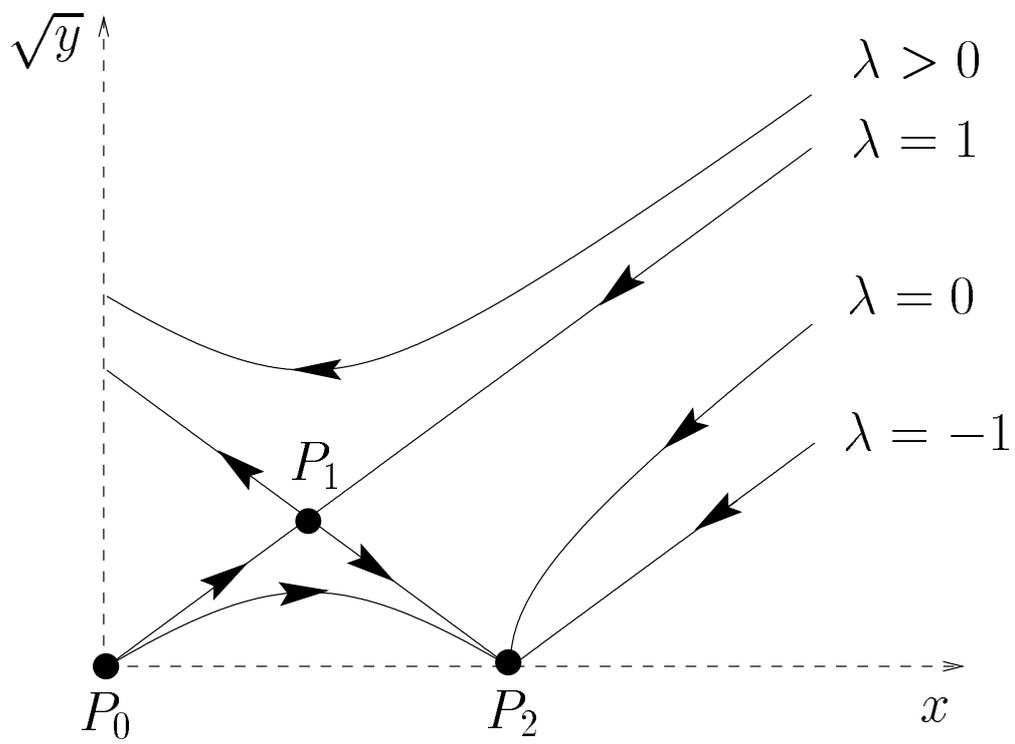
Figure 6: The flow pattern corresponding to eq. (4.4) for the limiting case $ \sigma = 3 $ 
(note that $F'(r=\infty) = 0 $ for $ \sigma = 3 $).

\end{document}